\documentclass[9pt, journal]{IEEEtran}
\usepackage{amsmath,amsfonts}
\usepackage{algorithmic}
\usepackage{orcidlink}
\usepackage{algorithm}
\usepackage{array}
\usepackage[caption=false,font=normalsize,labelfont=sf,textfont=sf]{subfig}
\usepackage{textcomp}
\usepackage{stfloats}
\usepackage{url}
\usepackage{amsmath}
\usepackage{amsthm}
\usepackage{tabularx}

\usepackage{verbatim}
\usepackage{graphicx}
\usepackage{cite}
\usepackage{bm}
\begin{document}

 \title{SLICE-CiM: \textbf{S}elective \textbf{L}ayer-wise \textbf{I}nference \textbf{C}ontrol \textbf{E}ngine with Confidence-Based Early Exit for SRAM based Digital \textbf{C}ompute-\textbf{i}n-\textbf{M}emory}

\author{Pratham Sharma\IEEEauthorrefmark{1}, Mohd Faisal Khan\IEEEauthorrefmark{1}, and Santosh Kumar Vishvakarma\IEEEauthorrefmark{1},  (Senior Member, IEEE)\\
\IEEEauthorblockA{\IEEEauthorrefmark{1}\textit{Department of Electrical Engineering}, \textit{Indian Institute of Technology Indore}, Indore, India}
\thanks{Manuscript received 20XX; revised  20XX.}
\thanks{This work was supported by the Special Manpower Development Program for Chip to Start-Up (SMDP-C2S), Ministry of Electronics and Information Technology (MeitY), Government of India, under Grant EE-9/2/21 - R\&D-E. (Corresponding author: Santosh Kumar Vishvakarma.)}
}

\markboth{IEEE TRANSACTIONS ON VERY LARGE SCALE INTEGRATION (VLSI) SYSTEMS,~Vol.~XX, No.~X, Month~20XX}%
{Shell \MakeLowercase{\textit{et al.}}: A Sample Article Using IEEEtran.cls for IEEE Journals}

\IEEEpubid{0000--0000~\copyright~2025 IEEE}

\maketitle
\begin{abstract}
Digital SRAM compute-in-memory (CIM) macros enable energy-efficient CNN inference by eliminating off-chip weight transfers; however, peripheral control logic introduces three inefficiencies: bit-slice over-activation at reduced precision (L1), unnecessary accumulation depth (L2), and idle precision-transition stalls (L3). To address these, we present four synthesizable peripheral blocks: a Flush-Merge Enable (FME) controller, a Confidence-Based Early-Exit (EE) module, a Precision Register with Decoder (PR/PD), and a Bit-Serial Cycle Controller (CC), co-designed to minimize switching activity, wordline overhead, and leakage. Implemented in TSMC 65-nm CMOS at 335\,MHz, the peripheral framework occupies 1769.760\,$\mu$m$^2$ post-synthesis and 1982.131\,$\mu$m$^2$ post-PnR. Gate-level VCD analysis reports 6.0706\,$\mu$W leakage and 288.2902\,$\mu$W (post-synthesis) / 322.8850\,$\mu$W (post-PnR) dynamic power, totaling 328.9556\,$\mu$W (0.2402\% of system power). The EE module dominates dynamic power (58.2\%), while WLG dominates area (58.1\%). System evaluation on MobileNetV2 (CIFAR-100) shows efficiency scaling from 31.35\,TOPS/W (8b/8b, 1.2\,V) to 454.91\,TOPS/W (2b/8b, 0.8\,V), achieving 13.33$\times$ improvement without bitcell modification or retraining. The framework targets CNN workloads exploiting sparsity, with Transformer extension as future work.
\end{abstract}

\begin{IEEEkeywords}
Compute-in-memory (CIM), digital SRAM, bit-serial MAC, mixed-precision inference, early-exit computation, precision switching, peripheral circuits, wordline gating, edge AI acceleration.
\end{IEEEkeywords}

\section{Introduction}
The rapid growth of deep neural network (DNN) workloads has increased the demand for energy-efficient real-time inference platforms. Conventional von Neumann architectures suffer from excessive data movement between memory and processing units, dominating latency and power. Compute-in-memory (CIM) architectures mitigate this bottleneck by performing multiply--accumulate (MAC) operations within memory arrays, reducing transfer overhead and improving efficiency \cite{ref1},\cite{ref2}. SRAM-based digital CIM (DCIM) macros are widely adopted due to CMOS compatibility, high speed, and scalability \cite{23},\cite{20}. Recent designs improve performance using programmable macros, bit-serial arithmetic, reconfigurable datapaths, and mixed-precision computation \cite{18},\cite{17},\cite{22},\cite{24}, while algorithm-level techniques such as mixed-precision quantization and hardware-aware training further enhance efficiency \cite{ref14},\cite{ref15},\cite{ref16}. Dynamic inference methods (e.g., early-exit) additionally reduce redundant computation. Despite these advances, prior works largely optimize array- and algorithm-level aspects independently, leaving peripheral control inefficiencies unresolved. Even with optimized arrays, peripherals introduce three key inefficiencies: (L1) bit-slice over-activation from imprecise wordline control, (L2) excess accumulation beyond convergence, and (L3) stalls during precision transitions. These stem from the lack of coordinated, precision- and confidence-aware control at the periphery. Although partially addressed via sparsity-aware and mixed-precision techniques \cite{ref16},\cite{17},\cite{22},\cite{24}, no prior work resolves all three jointly. This work presents an SRAM-based DCIM macro with a unified peripheral control framework integrating precision-aware slice activation, confidence-based early exit, and zero-stall precision transitions. By coupling control with workload dynamics, the design eliminates redundant activations, reduces accumulation overhead, and avoids pipeline stalls, significantly improving energy efficiency and throughput for neural inference workloads.

The main contributions of this work are:
\begin{enumerate}
\item A flush--merge enable controller that merges
accumulator flush, counter reset, precision update,
and wordline reconfiguration into the final
compute cycle, eliminating mode-transition stalls.
\item A registered-output early-exit mechanism
that terminates bit-serial accumulation after
confidence convergence, removing redundant
accumulation depth.
\item A precision register and decoder that
selectively activates bit-slices using a
training-free sensitivity metric, eliminating
slice over-activation.
\item A deterministic cycle controller that
supports safe early termination with a formally
bounded error constraint.
\end{enumerate}

The remainder of this paper is organized as follows. Section~II describes the baseline SRAM array and each proposed peripheral module from a transistor-level perspective. Section~III derives the formal early-exit error bound. Section~IV presents the training-free mixed-precision assignment. Section~V reports experimental setup and results, power analysis, and system-level evaluation. Section~VI concludes.

\section{Proposed DCIM Architecture}
\subsection{SRAM DCIM Array Specification and design methodology}
\begin{figure*}
    \centering
    \includegraphics[width=\linewidth]{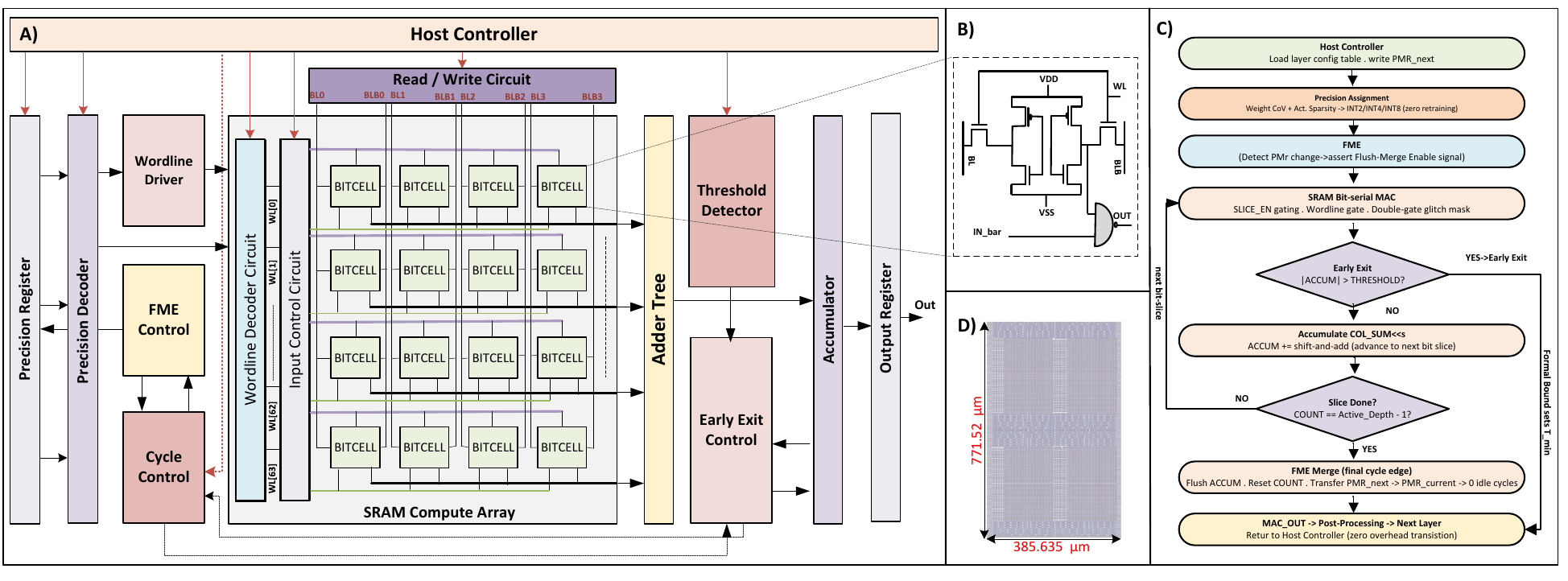}
    \caption{A) Proposed SLICE-CIM architecture, B) Conventional 10T NOR SRAM DCIM cell, C) SLICE-CIM operation, D) Layout of the 32~Kb SRAM CIM macro.}
    \label{fig:Arch}
\end{figure*}

The digital CIM array is based on a 10T NOR SRAM bitcell (Fig.~\ref{fig:Arch}(B)), selected for rail-to-rail compute bitline swing, strong hold stability under repeated wordline activation, and intrinsic CIM capability. The stored weight is non-destructively tapped from the $\overline{Q}$ node and combined with the input activation at the NOR gate, enabling in-situ 1-bit~$\times$~1-bit multiplication without disturbing the stored state. The array comprises eight independent $64\times64$ banks (32~Kb total), each storing one weight bit-slice. INT8 MAC is realized via bit-serial operation by activating one bank per cycle in MSB-first order, with all banks sharing the proposed six-block RTL peripheral fabric. The full-custom flow comprises six steps: (1) schematic design and transistor sizing in Cadence Virtuoso, with M1--M6 optimized for write margin and stability, and M7--M10 sized to sustain compute bitline discharge while preserving $\overline{Q}$ noise margins at 335\,MHz; (2) manual layout in Virtuoso Layout Suite following TSMC 65-nm GP rules, placing the NOR compute path on local interconnect to minimize parasitics; a verified $64\times64$ array is replicated across eight banks with shared decoders; (3) DRC sign-off using Mentor Calibre with zero violations; (4) LVS verification of full 32~Kb schematic-layout consistency; (5) parasitic extraction using Calibre; and (6) post-layout Spectre simulations across TT/1.2\,V/27$^\circ$C, SS/1.08\,V/125$^\circ$C, and FF/1.32\,V/$-40^\circ$C, confirming correct 1-bit multiplication and stable $\overline{Q}$ read-tap at 335\,MHz. The complete macro occupies $\sim$0.3\,mm$^2$, including a 0.01516\,mm$^2$ adder tree. Fig.~\ref{fig:Arch}(D) shows the SRAM DCIM layout with peripherals and adder tree. The array interfaces with control logic via $WL\_EN[7:0]$, $ACT[63:0]$, and $ADDR[5:0]$. A column adder tree with shift-and-add accumulation converts slice-wise partial sums $COL\_SUM[s]$ to the final MAC output. MSB-first ordering is enforced, where slice $k=0$ has weight $2^{(B-1)}$ and $k=B-1$ has weight $2^{0}$. A bit-serial MAC operation for a $B$-bit weight requires $B$ clock cycles, and the accumulator updates as:
\begin{equation}
  \texttt{ACCUM} \leftarrow \texttt{ACCUM}
    + \texttt{COL\_SUM}[k] \times 2^{B-1-k}
  \label{eq:accum}
\end{equation}

where the scale factor $2^{(B-1-k)}$ is realized as a fixed left-shift by $(B-1-k)$ positions in the adder-tree output path. The MSB-first ordering is a prerequisite for the confidence-based early-exit module, as it ensures each successive slice contributes at most half the magnitude of the preceding slice, bounding the tail error of early termination as derived in Section~III.
% \usepackage{graphicx}

\subsection{Precision Register (PR) and Precision Decoder (PD)}
The Precision Register and Precision Decoder (PR/PD) module resolves L1 by enabling runtime selection of the active wordline subset. A 2-bit Precision Mode Register (PMR) stores the current layer's precision mode. A fully combinational decoder maps PMR to the 8-bit wordline-enable vector $WL\_EN[7:0]$ and the 4-bit Active\_Depth value. Each bit of $WL\_EN[7:0]$ drives one wordline driver cell. When $WL\_EN[s]=0$ for $s \geq$ Active\_Depth, the corresponding the corresponding wordline driver holds those drivers static, eliminating all dynamic switching in inactive slices and resolving L1. At reset, both $PMR\_current$ and $PMR\_next$ initialize to 2'b01 (INT4 default), suppressing spurious FME assertion during the first inference layer. The PR/PD also maintains a shadow register $PMR\_next$ that stores the precision mode of the upcoming layer, loaded by the host controller one layer in advance. At reset ($rst\_n = 0$), both $PMR\_current$ and $PMR\_next$ are initialized to 2'b01 (INT4 default), ensuring:
\begin{equation}
\small
PRECISION\_CHANGE = PMR\_next \oplus PMR\_current = 0
\label{eq_prpd_reset}
\end{equation}
at startup and suppressing any spurious FME assertion during the first inference layer. PR comprises 7 standard cells and occupies 51.84 µm² post-synthesis (58.06 µm² post-PnR), representing 2.9\% of total peripheral area. PD uses 4 standard cells for 8.64 µm² (9.68 µm² post-PnR), 0.5\% of peripheral total.

\subsection{Flush-Merge Enable (FME) Controller}

The FME controller resolves L3 by parallelizing all four mode-transition operations into a single clock cycle at the layer boundary. A precision-change flag is computed combinationally as a 2-bit bitwise XOR of $\text{PMR}_{\text{next}}$ and $\text{PMR}_{\text{current}}$, followed by a 2-input OR gate. The FME pre-trigger is asserted during the penultimate compute cycle:
\begin{equation}
\begin{aligned}
\mathrm{FME} &= (\mathrm{COUNT} = \mathrm{Active\_Depth} - 2) \land \\
             &\mathrm{PRECISION\_CHANGE} \land (\mathrm{Active\_Depth} \geq 2)
\end{aligned}
\end{equation}
Upon assertion of $\text{FME}_{\text{final}}$, four register updates are committed within a single rising clock edge: the accumulator output register latches the current $\text{ACCUM}$ value; the accumulator state is reset; the cycle counter $\text{COUNT}$ resets to zero; and $\text{PMR}_{\text{current}}$ latches $\text{PMR}_{\text{next}}$. This eliminates all L3 idle stalls without pipeline penalty. The FME module uses 17 standard cells, occupies 35.280\,$\mu$m$^2$ post-synthesis (39.514\,$\mu$m$^2$ post-PnR), and accounts for 2.0\% of the total peripheral area.
\subsection{Cycle Controller (CC)}
The Cycle Controller governs the timing of all bit-serial operations and coordinates signals across the four peripheral modules. It maintains a 3-bit synchronous up-counter ($COUNT[2:0]$), generates the $DONE$ and $normal\_done$ completion signals, asserts $accum\_en$ for valid compute cycles, and enforces the safe-exit constraint:
\begin{equation}
\small
EX_{gated}
=
EX
\land
(COUNT \geq k_{min})
\label{eq_ex_gated}
\end{equation}
where $k_{min}$ = $\lfloor B/2 \rfloor$ is the minimum safe slice index derived in Section IV. It comprises 39 standard cells, occupies 109.80 µm² post-synthesis (122.98 µm² post-PnR), and represents 6.2\% of total peripheral area.

\subsection{Confidence-Based Early-Exit Module}

The Early-Exit (EE) module resolves L2 by dynamically terminating bit-serial accumulation once the intermediate MAC result reaches sufficient confidence. At each clock cycle $k$, the magnitude of the running accumulation is evaluated using two's-complement arithmetic. The negation is implemented via a 16-bit bitwise-NOT network followed by a 16-bit ripple-carry incrementer, with two pipeline stages inserted to meet the 335\,MHz timing constraint. The early-exit condition is:
\begin{equation}
\mathrm{EX} = \mathrm{accum\_valid} \land (|\mathrm{ACCUM}| > T)
\end{equation}
where $\mathrm{accum\_valid}$ is asserted only when $k \geq k_{\min} = \lfloor B/2 \rfloor$, enforcing the safe-exit region. The threshold $T$ is a 16-bit unsigned value in Q8.8 fixed-point format, loaded from a per-layer calibration table. This block is the largest peripheral module, comprising 154 standard cells and occupying 536.040\,$\mu$m$^2$ post-synthesis (600.365\,$\mu$m$^2$ post-PnR), accounting for 30.3\% of the total peripheral area. It also dominates dynamic power consumption (167.6591\,$\mu$W) due to the 16-bit magnitude computation and comparator pipeline.

\begin{figure}[!t]
    \centering
    \includegraphics[width=\linewidth,height=0.20\textheight,keepaspectratio]{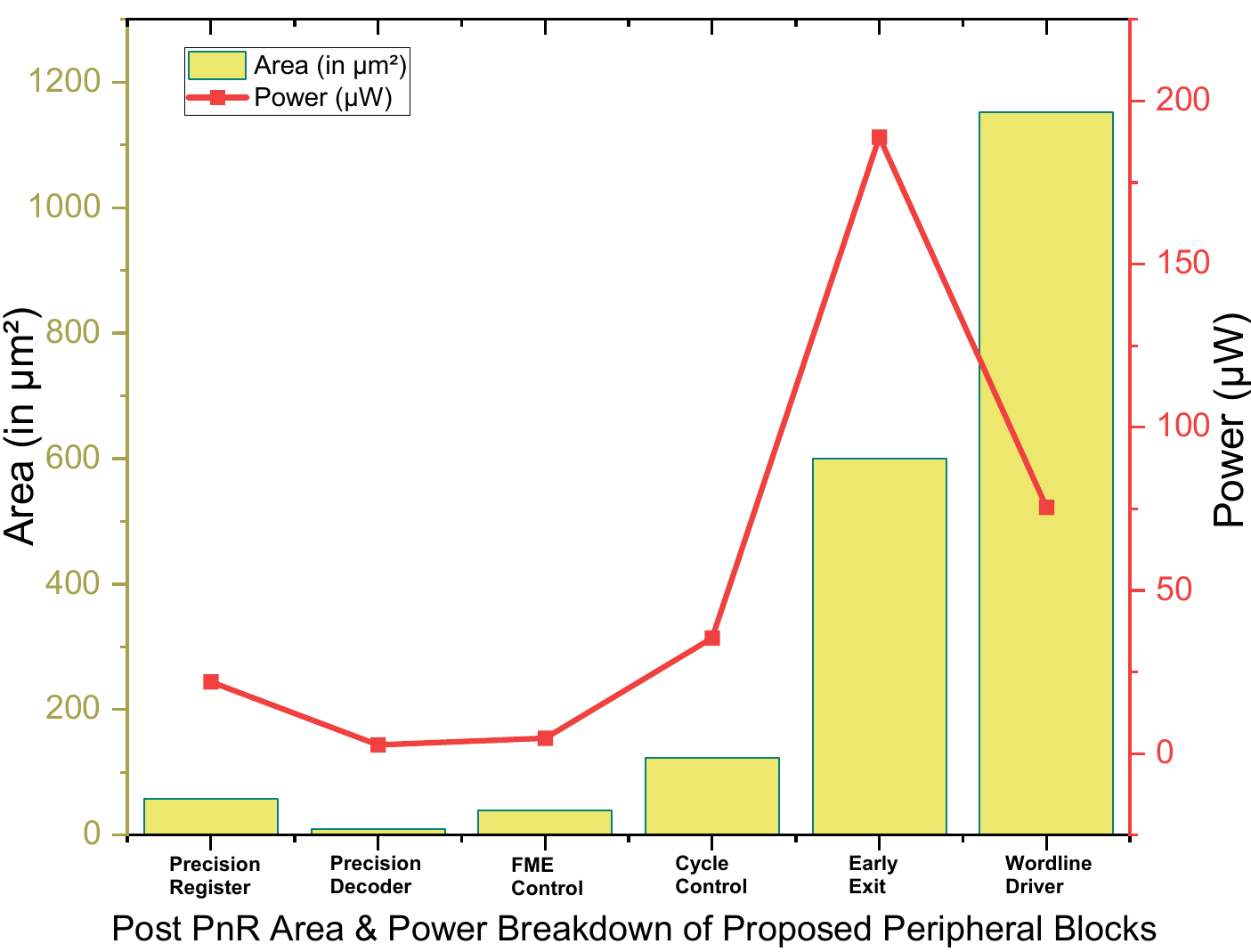}
    \caption{  Post PnR area and power breakdown of proposed peripherals}
    \label{fig:placeholder}
\end{figure}

\subsection{Wordline Driver}
The Wordline Driver interfaces the RTL peripheral fabric with the 10T NOR CIM array. It implements 8-slice $\times$ 64-column gating: for each slice $s$, the enable signal $\mathrm{WL\_EN}[s]$ is broadcast to all 64 column drivers. When $\mathrm{WL\_EN}[s]=0$, a NAND2 gate holds the driver input high, keeping access transistors (M5--M6) off and preventing wordline assertion. This disconnects the storage node from the NOR compute path, suppressing compute bitline discharge and adder-tree switching for inactive slices. This is the largest peripheral block, comprising 264 standard cells and occupying 1028.160\,$\mu$m$^2$ post-synthesis (1151.539\,$\mu$m$^2$ post-PnR), accounting for 58.1\% of total peripheral area. It also shows the highest switching power (46.6360\,$\mu$W), due to extensive activity across 64 columns. Fig.~\ref{fig:placeholder} presents the post-PnR area and power breakdown of the proposed peripheral blocks and \ref{fig:Arch}(C) shows the total operation flow of the work.
\section{ Formal Early-Exit Error Bound \& Per-Layer Threshold Calibration  }

\subsection{Formal Early-Exit Error Bound }

\textbf{Theorem 1:}
Let $w = [w_0, w_1, \ldots, w_{N-1}]$ be a weight vector with bounded magnitude $|w_i| \leq W_{max}$, and $x = [x_0, x_1, \ldots, x_{N-1}]$ be an activation vector with $|x_i| \leq 1$ (satisfied post-ReLU in CNN layers). Let $y_{full}$ denote the full-depth MAC result and $y_{EE}$ denote the early-exit output at slice index $k^*$. Then:
\begin{equation}
|y_{EE} - y_{full}|
\leq
W_{max}
\times
N
\times
2^{-(k^*+1)}
\times
\left(
2 - 2^{-(B-k^*-1)}
\right)
\label{eq_theorem1}
\end{equation}
\textit{Proof:}
Under MSB-first ordering, slice $k$ carries positional value $2^{(B-1-k)}$. When early exit occurs after slice $k^*$, the remaining slice contribution is bounded by evaluating the geometric sum:
\[
\sum_{j=k^*+1}^{B-1}
2^{(B-1-j)}
\]
\[
2^{-(k^*+1)}
\times
\left(
2 - 2^{-(B-k^*-1)}
\right)
\]
Summing across $N$ dot-product terms with $|x_i| \leq 1$ yields Equation~(\ref{eq_theorem1}).\\
\textbf{Corollary 1 (Safe Exit Region):} When exit occurs after at least half the precision depth:
\begin{equation}
k^* \geq \lfloor B/2 \rfloor
\end{equation}
\begin{equation}
|y_{EE} - y_{full}|
\leq
W_{max}
\times
N
\times
2^{-B/2}
\label{eq_corollary1}
\end{equation}
\textbf{Corollary 2 (Threshold Constraint):}
A minimum safe threshold is defined as:
\begin{equation}
T
>
W_{max}
\times
N
\times
2^{-B/2}
\label{eq_corollary2}
\end{equation}

ensuring the early-exit comparator fires only after the accumulated magnitude exceeds the maximum possible tail contribution.
\subsection{Per-Layer Threshold Calibration}

The confidence threshold $T$ for each layer is determined via an offline forward-pass calibration using 1,000 randomly selected training images (no test-set leakage).

A threshold sweep over 32 logarithmically spaced values within $[T_{min}, T_{max}]$ selects the value maximizing the Early-Exit Efficiency Factor (EEEF):

\begin{equation}
EEEF
=
\mathbb{E}
\left[
\frac{k^*}{Active\_Depth}
\right]
\label{eq_eeef}
\end{equation}
\begin{equation}
T_{min}
=
W_{max}
\times
N
\times
2^{-B/2}
\end{equation}

(from Corollary~2), and $T_{max}$ is defined as the 95th percentile of observed accumulator magnitudes at the final slice for that layer. Calibrated thresholds are stored as 16-bit Q8.8 fixed-point values in a per-layer lookup table. The threshold quantization step of:
\begin{equation}
2^{-8}
\approx
0.0039
\end{equation}
introduces a rounding error negligible compared to the plateau width of approximately 0.30 in normalized threshold units.\ref{tab_ablation_precision} shows the ablation study on mixed precision coefficient senstivity.

\section{Training-Free Mixed-Precision Assignment}
\begin{table}[!t]
\renewcommand{\arraystretch}{1.15}
\caption{Ablation Study: Mixed-Precision Coefficient Sensitivity}
\label{tab_ablation_precision}
\centering

\begin{tabular}{|c| c| c| c| c|}
\hline
$\alpha$ &
$\beta$ &
\textbf{TOPS/W} &
\textbf{Acc. (\%)} &
\textbf{Avg Prec. (bits)}
\\
\hline

0.5 & 0.5 & 37.91 & 69.42 & 4.58
\\
\hline
\textbf{0.6} & \textbf{0.4} & \textbf{41.63} & \textbf{68.70} & \textbf{4.31}
\\
\hline
0.7 & 0.3 & 39.40 & 68.12 & 4.19
\\
\hline
0.4 & 0.6 & 35.68 & 70.10 & 4.75
\\

\hline
\end{tabular}
\end{table}

\begin{figure}[!b]
    \centering
    \includegraphics[width=\linewidth,height=0.15\textheight,keepaspectratio]{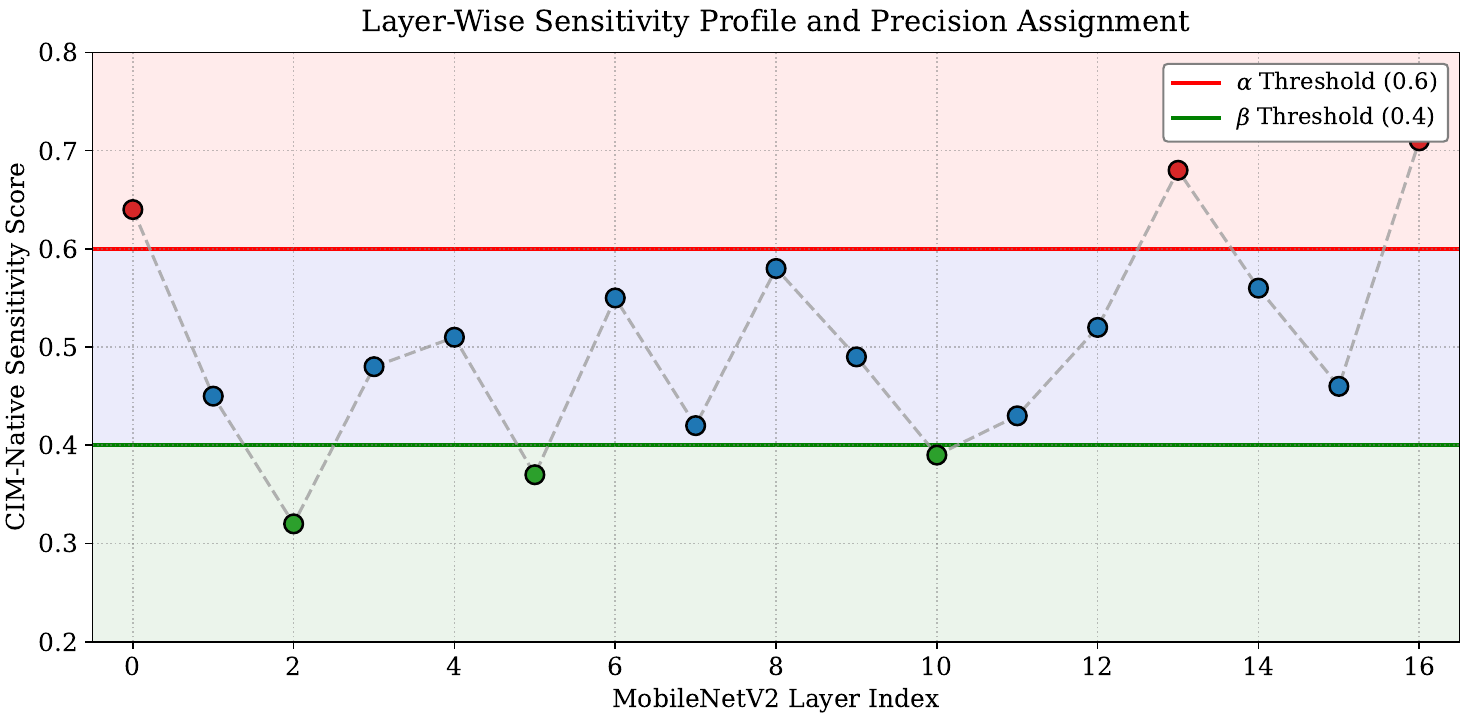}
    \caption{ Layer-wise sensitivity profile and precision assignment for MobileNetV2.}
    \label{fig:senstivity}
\end{figure}
Conventional mixed-precision methods such as HAWQ~\cite{ref14} require Hessian trace computation and quantization-aware retraining, while HAQ~\cite{ref15} relies on reinforcement learning with hardware-in-the-loop latency feedback. These approaches incur significant calibration overhead and are unsuitable for resource-constrained edge deployment. We propose a CIM-native sensitivity metric that maps directly to per-slice switching activity in the 65-nm SRAM array, enabling precision assignment through a single forward calibration pass without gradient computation or weight updates. For each layer $l$, sensitivity is quantified using two observable statistics. Weight variability is measured using the coefficient of variation (CoV):
\begin{equation}
CoV(l)
=
\frac{\sigma_w}{\mu_{|w|}}
\label{eq_cov}
\end{equation}
Activation sparsity after ReLU is:
\begin{equation}
\rho(l)
=
\frac{\text{Number of zero activations}}
{\text{Total activations}}
\label{eq_sparsity}
\end{equation}
The unified sensitivity score is:
\begin{equation}
S(l)
=
0.6
\times
CoV(l)
+
0.4
\times
(1 - \rho(l))
\label{eq_sensitivity}
\end{equation}
\begin{table}[!t]
\caption{Comparison of Proposed CIM-Native Precision Assignment With Existing Mixed-Precision Methods}
\label{tab:mp_compare}
\centering

\setlength{\tabcolsep}{1.25pt}
\renewcommand{\arraystretch}{1.15}

\begin{tabular}{lcccccc}
\hline

\textbf{Method} &
\textbf{Metric} &
\textbf{Retrain} &
\textbf{HW Model} &
\textbf{CIM} &
\textbf{Calib.} &
\textbf{Avg Bits} \\

\hline

EdMIPS~\cite{ref16} &
FLOP comp. &
Partial &
FLOPs &
No &
Low &
5.50 \\

HAWQ~\cite{ref14} &
Hessian trace &
Yes &
FLOP &
No &
Full QAT &
2.46 \\

HAQ~\cite{ref15} &
RL reward &
Yes &
Latency LUT &
No &
High &
6.00 \\

\textbf{This Work} &
CoV + sparsity &
\textbf{No} &
\textbf{CIM slice} &
\textbf{Yes} &
\textbf{1-pass} &
\textbf{4.31} \\

\hline
\end{tabular}

\vspace{3pt}   % <-- This fixes spacing

\footnotesize{
QAT: Quantization-Aware Training; 
RL: Reinforcement Learning; 
CoV: Coefficient of Variation.
}

\end{table}
Precision modes are assigned using thresholds $S_H$ and $S_L$:
\begin{equation}
PMR(l)
=
\begin{cases}
INT8, & S(l) > S_H \\
INT4, & S_L \leq S(l) \leq S_H \\
INT2, & S(l) < S_L
\end{cases}
\label{eq_precision_rule}
\end{equation}

For MobileNetV2 on CIFAR-100, $S_H = 0.65$ and $S_L = 0.35$, mapping 3 layers to INT8, 11 to INT4, and 3 to INT2, yielding a weighted-average precision of 4.31 bits. Fig.\ref{fig:senstivity} shows Layer-wise sensitivity and precision assignment for MobileNetV2. The horizontal lines indicate optimal thresholds ($\alpha = 0.6$, $\beta = 0.4$), yielding a 3/11/3 split: highly sensitive layers use INT8 (green), moderate layers INT4 (blue), and robust layers INT2 (red), maximizing bit-serial efficiency while preserving accuracy.

\begin{table*}[!t]
\renewcommand{\arraystretch}{1.4}
\caption{Comprehensive Comparison with State-of-the-Art Works}
\label{tab:SOTA Comp}
\centering
\scriptsize
\resizebox{\textwidth}{!}{
\begin{tabular}{|c|c|c|c|c|c|c|c|c|c|c|}
\hline
\textbf{Parameters} & \textbf{TCAS-I'25 \cite{23}} & \textbf{TCAS-II'22} \cite{24} & \textbf{JSSC'22} \cite{17} & \textbf{TCAS-I'23} \cite{18} & \textbf{JSSC'21} \cite{19} & \textbf{TCAS-I'24} \cite{20} & \textbf{TCAS-I'23} \cite{21} & \textbf{JSSC'23} \cite{22} & \textbf{ISSCC'21} \cite{ref2} & \textbf{Proposed} \\
\hline
Tech. (nm) & 65 & 65 & 65 & 65 & 65 & 55 & 28 & 28 & 22 & 65 \\
\hline
Operation & Digital & Analog & Analog & Digital & Digital & Digital & Digital & Digital & Digital & Digital \\
\hline
Cell-type & 8.75T & 6T & 6T & 7T & 54T & 8T & 6.5T & 10T & 6T & 10T \\
\hline
VDD (V) & 1 & 0.9--1.2 & 0.9--1.05 & 0.8--1.1 & 0.6--0.8 & 1.2 & 0.6--1.1 & 0.6--0.9 & 0.72 & 0.8-1.2 \\
\hline
Macro (Kb) & 64 & 72+1 & 16 & 80 & 16 & 64 & 16 & 128 & 64 & 32 \\
\hline
Frequency (MHz) & 400 & 100 & 100 & 200 & 138 & 200 & 360 & 216 & 55.56 & 335 \\
\hline
Area (mm$^2$) & 2.35 & 2.79 & 9 & 0.473 & 1 & 2.8/3.91* & 0.59/3.43* & 8.97/48.34* & 0.202 & 0.346 \\
\hline
Input (bit) & 1--8 & 4 & 2/4/6/8 & 1--16 & 1--16 & 4/8/12/16 & 1--8 & 4/8 & 1--8 & 8 \\
\hline
Weight (bit) & 1--8 & 4 & 4/8 & 1--16 & 1--16 & 4/8/12/16 & 8 & 2/4/8 & 4/8/12/16 & 2/4/8 \\
\hline
Model & NA & VGG-11 & VGG-16 & ResNet50 & Lenet-5 & ResNet-18 & NA & ResNet20 & LeNet-5 & MobileNetV2 \\
\hline
Accuracy (\%) & NA & 87.88 & 99.19 & 93 & 99.2 & 94.5 & NA & 90.8 & 98.3 & 68.70 \\
\hline
Artificial Sparsity (\%) & No & Yes (90.02\%) & Yes (72.5\%) & Yes (90\%) & No & No & Yes (10\%) & Yes & No & No \\
\hline
Energy Efficiency (TOPS/W) & 13.84/6.79*(8/8) & 33.14/5.485*(8/8) & 5.78(4/8) & 315.07/4.92*(8/8) & 6.87/1.21*(8/8) & 16.6/9.83*(8/8) & 60.4/4.59*(8/8) & 5.99/0.32*(8/8) & 24.7/2.13*(8/8) & 90.58 (4/8), 31.35 (8/8) \\
\hline

\end{tabular}
}
\vspace{2pt}
\footnotesize * Scaled for macro-size, Tech-node and Voltage 1.2V .
\end{table*}

\section{Results and Discussion}

\subsection{Experimental Setup}
The proposed system comprises two layers: (1) a full-custom 10T NOR SRAM CIM macro and (2) an RTL peripheral control framework synthesized in Cadence Genus, integrated via a full-custom-to-digital ASIC flow. The CIM cell was designed in Cadence Virtuoso under TSMC 65-nm GP rules, combining a 6T storage core with a 4T NOR compute extension, with routing optimized to minimize parasitics. A $64\times64$ array was implemented, verified using Mentor Calibre (DRC/LVS clean), and post-layout parasitics were extracted using Calibre xRC (M1–M4).
Post-PEX Spectre simulations across TT/1.2\,V/27$^\circ$C, SS/1.08\,V/125$^\circ$C, and FF/1.32\,V/$-40^\circ$C confirmed correct 1-bit NOR multiplication and non-destructive read at 335\,MHz. The validated array was exported as a GDSII hard macro, with .lef generated via Cadence Abstract Generator and .lib obtained from post-PEX characterization (timing, capacitance, power across PVT). A behavioral Verilog model with back-annotated delays was also created. Peripheral modules were implemented in synthesizable Verilog and compiled using Cadence Genus with the TSMC 65-nm library. The CIM macro (.lib/.lef) was included during synthesis for accurate timing and loading, targeting 335\,MHz (2.985\,ns) using a segmented wireload model. Power was estimated via gate-level VCD flow: MobileNetV2 inference on 500 CIFAR-100 images was simulated in Cadence Xcelium to generate switching activity, which was back-annotated into Genus Power Analysis with interconnect RC and macro power models to extract instance-level dynamic and leakage power, with dynamic power derived from VCD-based toggle activity.

\subsection{Results Calculation }
Table~I summarizes the synthesis area. The six-block peripheral framework occupies 1769.760\,$\mu\mathrm{m}^2$ post-synthesis, corresponding to 0.380\% overhead relative to the 10T NOR SRAM CIM macro (19{,}880\,$\mu\mathrm{m}^2$). Post-PnR, the area increases to 1982.131\,$\mu\mathrm{m}^2$ (0.425\% vs.\ 22{,}265.60\,$\mu\mathrm{m}^2$). The wordline driver dominates area at 1028.160\,$\mu\mathrm{m}^2$ (58.1\%), followed by the early-exit module at 536.040\,$\mu\mathrm{m}^2$ (30.3\%). Peripheral dynamic power is 288.2902\,$\mu\mathrm{W}$ post-synthesis (322.8850\,$\mu\mathrm{W}$ post-PnR), with 6.0706\,$\mu\mathrm{W}$ leakage. The early-exit block dominates dynamic power at 167.6591\,$\mu\mathrm{W}$ (58.2\%), while the wordline driver has the highest toggle component (46.6360\,$\mu\mathrm{W}$). Total peripheral power is 294.3608\,$\mu\mathrm{W}$ (0.2402\% of system power, 122{,}532\,$\mu\mathrm{W}$), whereas the CIM array consumes 118{,}799\,$\mu\mathrm{W}$ (96.95\%), with 117{,}661.56\,$\mu\mathrm{W}$ dynamic. Peripheral dynamic power remains nearly constant (288.2902\,$\mu\mathrm{W}$) as the array scales from $64\times64$ to $256\times256$, due to fixed-width control logic, while array power scales with column count. This results in monotonic improvement in system-level TOPS/W with increasing array size. Table~\ref{tab:mp_compare} compares the proposed design with state-of-the-art mixed-precision methods, and Table~\ref{tab:SOTA Comp} provides a comprehensive comparison with prior CIM works.

\begin{figure}[t]
    \centering

   \includegraphics[width=\linewidth,height=0.25\textheight,keepaspectratio]{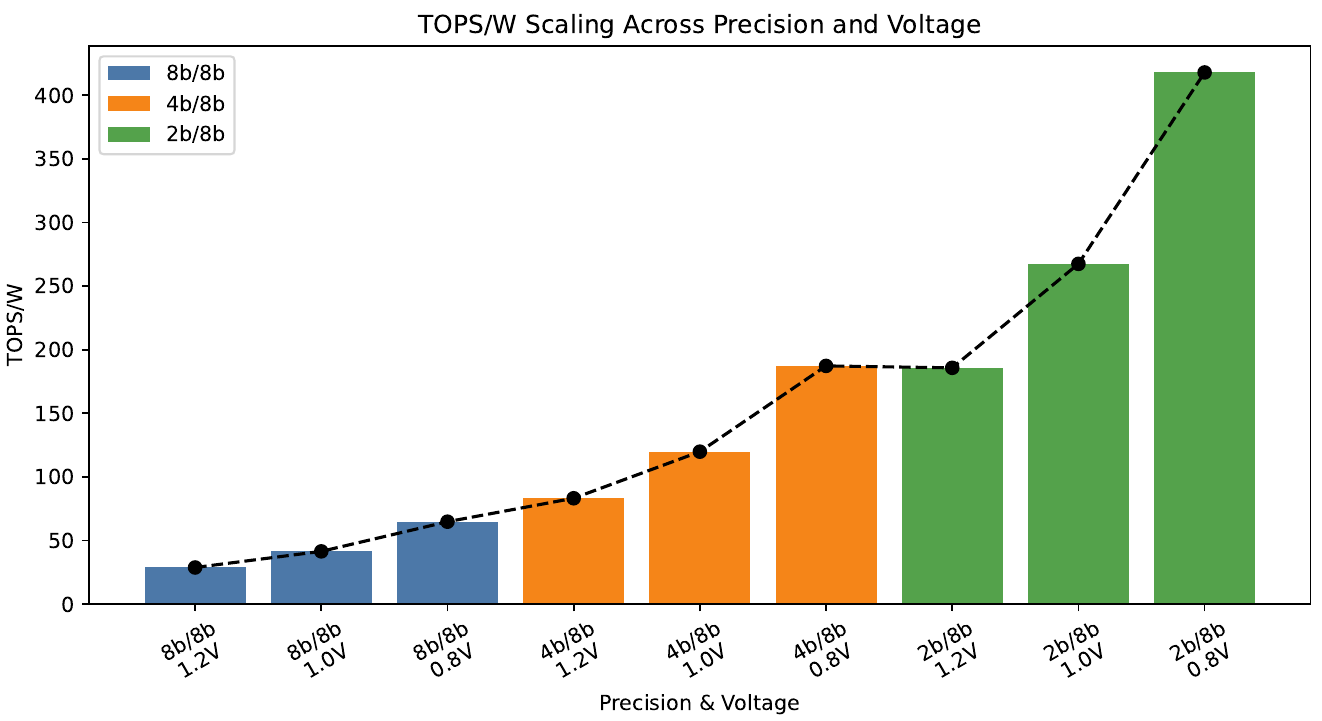}
    \caption{Energy efficiency analysis at different precisions and voltages.}
    \label{fig:perf_combined}
\end{figure}

\subsection{System-Level Accuracy and Efficiency Scaling}

To evaluate application-level performance, MobileNetV2 inference was performed on CIFAR-100 (10{,}000 images). Using mixed-precision thresholds ($\alpha=0.6$, $\beta=0.4$), 3 layers map to INT8, 11 to INT4, and 3 to INT2, yielding an average precision of 4.31 bits with 68.70\% Top-1 accuracy (vs.\ 70.10\% baseline). Under this workload, the macro achieves 41.63\,TOPS/W at 1.2\,V and 335\,MHz. A comprehensive TOPS/W sweep across precision and supply voltage was also performed. At 8b/8b and 1.2\,V, the macro achieves 31.35\,TOPS/W. Reducing precision to 4b/8b increases sparsity and enhances early-exit (EE) skipping and spatial gating, improving efficiency to 90.58\,TOPS/W (2.65$\times$). Further voltage scaling to 0.8\,V yields 187.25\,TOPS/W. Peak efficiency is achieved at 2b/8b and 0.8\,V, reaching 454.91\,TOPS/W (13.33$\times$ over the baseline), as shown in Fig.~\ref{fig:perf_combined}. These results demonstrate the synergistic benefits of precision scaling, sparsity-aware execution, and voltage scaling in maximizing system-level energy efficiency.

\section{Conclusion}
This work presents a precision-scalable peripheral control framework co-designed with a full-custom 10T NOR CIM macro in TSMC 65-nm CMOS. The 10T cell extends a standard 6T SRAM with four transistors to enable in-situ NOR computation, where the stored weight is non-destructively tapped from the $\overline{Q}$ node and combined with the input activation to realize a 1-bit multiplier. Six RTL peripheral blocks address bit-slice over-activation (L1), excess accumulation depth (L2), and precision-transition stalls (L3). Synthesized at 335\,MHz in Cadence Genus using a TSMC 65-nm library, the peripherals occupy 1769.760\,$\mu\mathrm{m}^2$ post-synthesis (1982.131\,$\mu\mathrm{m}^2$ post-PnR), corresponding to 0.380\% (0.425\% post-PnR) area overhead. Gate-level VCD analysis reports 288.2902\,$\mu\mathrm{W}$ dynamic and 6.0706\,$\mu\mathrm{W}$ leakage power, totaling 0.2402\% of system power. The design achieves 454.91\,TOPS/W at 2b/8b and 0.8\,V, providing a 13.33$\times$ improvement over the 8b/8b, 1.2\,V baseline (31.35\,TOPS/W) without retraining or bitcell modification. The framework is validated on CNN inference workloads exploiting ReLU-induced sparsity and bounded activations. Extension to Transformer models remains future work due to lower sparsity in GELU/SiLU activations, input-dependent attention distributions, and activation outliers violating the $|x_i|\leq1$ assumption, requiring adaptive thresholds and outlier-aware quantization.

\bibliographystyle{ieeetr}
\bibliography{bib}

\end{document}